# DARK MATTER AND STABLE BOUND STATES OF PRIMORDIAL BLACK HOLES

L. K. Chavda*
Physics Department
South Gujarat University
Udhna-Magdalla Road,
Surat – 395007, Gujarat, India.
E-mail : lk_chavda@rediffmail.com
And
Abhijit L. Chavda
49, Gandhi Society*,
City Light Road,
Via Parle Point.
Surat 395007, Gujarat, India.
E-mail : al_chavda@rediffmail.com.

ABSTRACT

We present three reasons for the formation of gravitational bound states of primordial black holes, called holeums, in the early universe. Using Newtonian Gravity and nonrelativistic quantum mechanics we find a purely quantum mechanical mass-dependent exclusion property for the nonoverlap of the constituent black holes in a holeum. This ensures that the holeum occupies space just like ordinary matter. A holeum emits only gravitational radiation whose spectrum is an exact analogue of that of a hydrogen atom. A part of this spectrum lies in the region accessible to the detectors being built. The holeums would form haloes around the galaxies and would be an important component of the dark matter in the universe today. They may also be the constituents of the invisible domain walls in the universe.

---

* Present address

I. INTRODUCTION

The recent International Conference on Gravitational Waves held in Rome has described the twenty first century as the century of gravitation. It has called upon the theoretical physicists to investigate the gravitational waves and to make testable predictions. In the Massachusetts Institute of Technology the Laser Interferometer Gravitational Observatory (LIGO),among others, has been set up for the detection of gravitational waves. This paper is an attempt to address the issue. The dark matter in the universe is inferred to be several times the observed one. Therefore it is natural to look for a cosmological solution to the problem of its origin. One may consider the vast amount of the primordial black holes produced during the early history of the universe. If they formed stable gravitational bound states then they may constitute a large component of the dark matter in the present day universe. Besides, the transitions among the energy levels of such purely gravitational bound states would result in the emission of gravitational radiation in exact analogy to the atoms radiating electromagnetic radiation under similar circumstances. However, an isolated black hole is subject to the evaporation of its mass due to the Hawking radiation caused by the vacuum fluctuations in its vicinity. This would seem to rule out the possibility of formation of their bound states. However, there are three mitigating factors in favour of the formation of the latter. (1) When the temperature of the bigbang universe is much greater than $T_b=mc^2/k_B$, where m is the mass of a black hole and $k_B$, the Boltzmann constant, the number density of such black holes is proportional to $(k_B T)^3$. For the temperatures of the early universe of interest to us in this paper this quantity is very large, lying between $10^{25}$ and $10^{57}$ $GeV^3$. This vitiates the condition of isolation of black holes necessary for the Hawking radiation. (2) Near the unification temperature, of the order of $10^{16}$ GeV, all the four fundamental interactions of nature are expected to have the same strength. In particular, the gravitational interaction would have a strength far greater than it has now. This vitiates another condition for the Hawking radiation, namely, a field-free space.(3)This means that the rate of gravitational interactions among the black holes would correspondingly be higher. And as long as it remains higher than the rate of expansion of the early universe, there would be sufficient number of black holes in very close vicinity to form bound states of black holes called holeums[1]. In other words, extremely high number density, vastly stronger gravity and an enormously larger rate of interactions are likely to lead to the formation of the holeums. It is instructive to note that whereas a free neutron decays, neutrons in the nucleosynthesis era of the early universe finding themselves in high number densities and subject to the strong interaction formed stable nuclei and never decay except those neutrons that are in heavy nuclei containing a large number of protons. This is very similar to the formation of stable holeums as bound states of unstable primordial black holes. Therefore we make the plausible Ansatz that holeums were produced in the early universe and explore its consequences in the following.

To solve the problem of the gravitational bound states we must quantise the gravitational field. This is an old open problem. The usual attempts to quantise the gravitational field lead to paradoxes and inconsistencies both in the nonrelativistic and the relativistic domains[2]. The recent attempts in the form of superstrings and supergravity theories are still in a flux. Today we face a situation similar to those faced by N.Bohr. and S. Glashow. Bohr wanted to derive the formula for the spectrum of the hydrogen atom. He seized the then newly emergent idea of quantisation and succeeded in his task. Similarly, in order to cancel divergent diagrams in the then putative unified theory of weak and electromagnetic interactions, Glashow proposed a new charm-quark which was discovered subsequently. Both attempts were successful but faulty. Bohr's was in conflict with the subsequently discovered uncertainty principle and was , therefore, discarded completely but in the process it catalysed the emergence of the new quantum theories of wave mechanics and matrix mechanics. Glashow's idea eventually became a part of the renormalised, unified electroweak theory. Thus, it is clear that existing rudimentary theories may give important insights into new phenomena and may catalyse new, complete theories. We, therefore, try in the following to see what light the existing theories throw on the problem of the dark matter in the universe.

But first we must address several issues: (a) The choice of the theory.(b)The classical or the quantum nature of the system (c) The finite size of the black holes (d)The form of the potential inside and outside the bound state and the Bound State Wisdom.(e) The Paradigm of the Truncated Hydrogen Atom.

(a)The Theory: The quantum mechanics of two neutral black holes is a problem which remains largely unstudied in General Relativity (GR) due to the lack of knowledge of an appropriate exact analytic solution to the relevant two-body problem. Progress has been made with the study of quantum mechanical *scattering* of exremally charged black holes[3-7], since an appropriate exact multicentre metric[8] is known in such cases. Our interest, however, is in purely gravitational bound states of neutral black holes. Thus, in the absence of an exact solution in GR we will proceed by an analysis of the analogous problem in Newtonian Gravity (NG). The latter and the GR both lead to paradoxes and inconsistencies when quantised[2].Thus, there is no guarantee that a quantised GR will give better results than a quantised NG. Therefore we choose to work with the latter and offer the following additional reasons for the choice[9]: (1) " It reveals the implicit simplicity of the cosmological equations …and offers insight into their physical nature".(Harrison 1965).(2) "It reveals all the essential features of relativistic cosmology without the mathematical complexity. It also makes the significance of the different terms apparent much more readily".(Bondy 1952).(3) " (It prepares) the way towards understanding relativistic cosmology….Not only is Newtonian theory mathematically simpler, it also leads to many results that are essentially the same as in relativity". (Sciama 1971).(4) " It is not only of pedagogical value but also of great heuristic value since (some problems) are too difficult to be taken into account within the framework of the general theory of relativity". (Zeldovich 1965).

In summary, we choose the quantised NG for lack of manifestly superior quantised alternative, simplicity, transparency and exact solvability of the problem under consideration.The region of applicability of NG is $r \gg R$, the Schwarzschild radius of the black hole.

(b)Quantum Nature: Next we must consider whether the holeums are classical systems or quantum ones. A criterion for this is the following : If the de Broglie wavelengths of the constituents of a bound state are comparable to the linear size of the latter, then the system is a quantum one. To this end, consider for simplicity, a black hole confined in a cubic box of side $l$. Applying the uncertainty principle, one can show that the de Broglie wave length of the black hole is $\lambda \cong 3^{-1/2} l$ which is comparable to $l$. Now the holeums of interest to us are of atomic and nuclear sizes. Hence we conclude that these holeums are quantum systems.

(c)The Finite Size of the Black Holes: When the radius of the bound state is far greater than that of the constituents we may regard the latter effectively as point particles. For example, the holeums of atomic and nuclear sizes have radii of $10^{-10}$ m and $10^{-15}$ m, respectively, whereas the black holes in this paper have radii less than $10^{-35}$ m .Hence they may be regarded as point particles. But when the mass of the black hole is less than but close to $m_P$, where the latter is the Planck mass, the size of the bound state is only a small factor times R. Here the black holes cannot be regarded as point particles. Not only that, we are now in the strong field regime where the NG is no longer valid. This is very similar to the deuteron problem: The neutron and the proton both have radii of about 0.8 Fermi whereas the radius of the deuteron is about 2.34 Fermi. Thus, the nucleons cannot be regarded as point particles. And yet they are so regarded in solving the Schrodinger equation and one gets reasonable values of the bound state parameters. This is because the bound state problem, in general, is not too restrictive of its parameters. Secondly, the bound state does not probe the region around $r = 0$.To probe the latter, one needs to resort to the high energy scattering. It is also interesting to note that asymptotic models quite often give very reasonable results outside their regions of validity. For example, the Bohr-Sommerfeld quantisation condition derived from the WKB method, valid for $n \gg 1$, is known to give exact results even for $n = 1$ for the case of hydrogen atom and reasonable ones for oscillator and other problems.

(d)The form of the potential inside and outside the bound state and the Bound State Wisdom: Consider two examples from nuclear and particle physics. In the deuteron problem some treatments take an infinitely repulsive hard core potential inside an attractive square well one while others neglect the very singular former one. And yet one gets acceptable values of the bound state parameters. In quarkonia, a linear plus a Coulomb, a logarithmic and a small positive power potential all give quantitatively acceptable description of the Charmonium and the Bottonium spectra. In the first case the potential is very singular, in the second case it is mildly so and in the third case it is nonsingular. And yet all of them give quantitatively acceptable fits. These examples show that as long as it leads to a bound state the acceptable potential can be drastically different within the bound state. We call this the Bound State Wisdom (BSW) and demonstrate its efficacy by applying it to the well-known hydrogen atom problem very relevant to us. This will create a benchmark useful to us.

(e) A Paradigm of the Truncated Hydrogen Atom (POTHA): We now extend the BSW to the third, atomic, layer of matter. We drastically modify or truncate the $r^{-1}$ potential of a hydrogen atom within and near the bound state while keeping its asymptotic behaviour the same. Since the same $r^{-1}$ asymptotic potential governs the formation of a holeum, this paradigm is of direct relevance to us in this paper. We carry out the analysis using dimensionless quantities so that our results will be scale-invariant when we apply them to the holeum. Consider the modified potential for the hydrogen atom.

$$V(r) = -V_1 \qquad 0 \leq r \leq r_0$$
$$\phantom{V(r)} = -e^2/r \qquad r_0 \leq r \leq \infty \qquad (1)$$

where
$$V_1 = e^2/r_0 \qquad (2)$$

The radial Schrodinger equation for s-states is given by
$$d^2u/dr^2 + (2\mu/\hbar^2)(E - V(r))u(r) = 0 \qquad (3)$$

For the region 1, $0 \leq r \leq r_0$, this reduces to
$$d^2u/dr^2 + (2\mu/\hbar^2)(V_1 - |E|)u(r) = 0 \qquad (4)$$

where $\mu$ is the reduced mass. Now we define some dimensionless quantities.

$$\lambda_1^2 = 2p(1-p\varepsilon) \qquad (5)$$
$$x = r/r_0 \qquad (6)$$
$$p = r_0/a \qquad (7)$$
$$a = \hbar^2/(\mu e^2) \qquad (8)$$
$$\varepsilon = |E|/V_0 \qquad (9)$$
$$V_0 = e^2/a \qquad (10)$$

In terms of these quantities we can rewrite Eq.(4) as
$$u''(x) + \lambda_1^2 u(x) = 0 \qquad (11)$$

The solution to this equation with $u(0) = 0$ is given by
$$u(x) = A \sin \lambda_1 x \qquad (12)$$

For the region 2, $r_0 \leq r \leq \infty$, or $1 \leq x < \infty$, we denote the wave function by $v(r)$. It satisfies the equation
$$d^2v/dr^2 + (2\mu/\hbar^2)(e^2/r - |E|) v(r) = 0 \qquad (13)$$

In terms of the quantities introduced above this reduces to
$$v''(x) = 2p(p\varepsilon - 1/x) v(x) \qquad (14)$$

The asymptotic solution to Eq.(14), as $x \to \infty$, is given by
$$v(x) \to \exp(-\lambda_2 x) \qquad (15)$$

where
$$\lambda_2^2 = 2p^2 \varepsilon \qquad (16)$$

In a rigorous variational treatment one would treat $\lambda_1$, $\lambda_2$ and $r_0$ as the three variational parameters and minimize the expectation value of the Hamiltonian with respect to them. But since our goal is modest we adopt the following approximate treatment designed to show the plausibility of obtaining the ball-park values. We consider only the 1s state and take its wave function

$$v(x) = (a + bx)\exp(-\lambda_2 x) \qquad (17)$$

The continuity of the wave functions at $x = 1$ gives us
$$A \sin \lambda_1 = (a + b) \exp(-\lambda_2) \qquad (18)$$

The continuity of the derivatives at $x = 1$ leads to
$$A \lambda_1 \cos\lambda_1 = (b - (a+b)\lambda_2) \exp(-\lambda_2) \qquad (19)$$

From the last two equations we get
$$\lambda_1 \cot\lambda_1 = -\lambda_2 + b/(a+b) \qquad (20)$$

Now from Eq.(17) we get
$$v''(1) = (-2b\lambda_2 + (a+b)\lambda_2^2) \exp(-\lambda_2) \qquad (21)$$

and from Eq.(14) we have

$$v''(1) = 2p(p\varepsilon - 1)(a + b) \exp(-\lambda_2) \qquad (22)$$

From Eqs.(20-22) we get

$$\lambda_2^2 - \lambda_1^2 + 2\lambda_1\lambda_2 \cot\lambda_1 = 0 \qquad (23)$$

This equation depends upon two dimensionless parameters p and ε. For a given value of one of them the value of the other is obtained by solving Eq.(23). Several results are presented in Table I. From the latter we see that the most probable radius $r_{max}$, obtained by maximising $x^2 v(x)^2$, varies between a/5 and 3.23 a where a is the first Bohr radius of the real hydrogen atom. The ground state energy of the truncated hydrogen atom varies between 60 % and 80 % of that of the real hydrogen atom. But it must be noted that the value ε =1/2 which corresponds to the ground state energy of the real hydrogen atom is inaccessible to the truncated one. But values quite close to it, about 80 % of it, are obtained for the latter. As a "first glance" solution this would have been a valuable contribution to the atomic physics.

In summary, we conclude that for holeums of atomic and nuclear sizes the NG rules both within and without the holeums and we will take the $r^{-1}$ potential for $0 \leq r \leq \infty$. For the strong field case this is not true near and inside the bound state. However, for r>>1 the potential is still $r^{-1}$ even in the strong field regime. But in view of the BSW and POTHA we will take the potential to be $r^{-1}$ everywhere. Since the results of the paradigm are scale–invariant we can expect results acceptable within a factor of several for the size of the holeum and a lot more accurate result for the energies and the frequencies of the holeums.

In this paper we consider black holes in the mass range $10^3$ GeV/$c^2$ to $10^{15}$ GeV/$c^2$. This means that we can treat a holeum as a nonrelativistic system just like the heavy quarkonia.

In section II, we consider a nonrelativistic quantised gravitational bound state : a holeum. We derive its energy and mass eigen values and an expression for its size. Section III deals with the gravitational radiation emitted by a holeum. In section IV we establish a connection between holeums and the dark matter in the universe. The haloes of holeums around mass concentrations may provide an explanation for the cosmological problem of the missing domain walls. Section V deals with the possibility of production of holeums in terrestrial and cosmic sources. Conclusions are presented in section VI.

II.   A STABLE HOLEUM.

For the sake of simplicity we consider two identical black holes of mass m. The gravitational potential between them is given by

$$V(r) = -m^2 G / r = -\alpha_g \hbar c / r \qquad (24)$$

where r is the distance between them. ℏ, c and G are the Planck's constant reduced by 2π, the speed of light in vacuum and Newton's universal gravitational constant, respectively. $\alpha_g$ is the gravitational analogue of the fine structure constant, given by

$$\alpha_g = m^2 G / \hbar c = m^2 / m_p^2 \qquad (25)$$

where

$$m_p = (\hbar c/G)^{1/2} \qquad (26)$$

is the Planck mass. The Schrodinger equation is exactly solvable for the $r^{-1}$ potential and the energy eigen values, formally identical with those of the hydrogen atom, are given by[10]

$$E_n = -\mu c^2 \alpha_g^2 / (2n^2) \qquad (27)$$

where n is the principal quantum number, n = 1,2,… ∞ and μ = m/2 is the reduced mass. In the following we will consider, for simplicity, only the l=0, s-states. The eigen function for an ns state is given by[10]

$$\Psi_{ns} = A_n L^1_{n-1}(t) e^{-t/2} \qquad (28)$$

where

$$t = 2\chi r \qquad (29)$$
$$\chi = \alpha_g^2 / (nR) \qquad (30)$$
$$R = 2mG/c^2 \qquad (31)$$

which is the Schwarzschild radius of the black hole. Here $L^m_n(x)$ is the associated Laguerre polynomial and

$$A_n^2 = 4\chi^3 / (n^2 (n!)^2) \tag{32}$$

The maxima of the probability density

$$g(r) = r^2 |\Psi_{ns}|^2 \tag{33}$$

give us the radii of the stable orbits. For the 1s state the radius of the orbit is given by
$$r_1 = R/\alpha_g^2 \tag{34}$$
For the 2s state there are two orbits with radii
$$r_{2\pm} = (3\pm\sqrt{5}) R/\alpha_g^2 \tag{35}$$
For m>>1 we have [11]
$$L^\alpha_m(x) \cong \pi^{-\frac{1}{2}}(m+\alpha)! \, x^{-\alpha/2-1/4} m^{\alpha/2-1/4} e^{x/2} \cos[2(mx)^{1/2} - \alpha\pi/2 - \pi/4] \tag{36}$$
Using this we can show that for n,n' >>1, the radii of the stable orbits are given by
$$r_n = n'^2 \pi^2 R/(8\alpha_g^2) \tag{37}$$
where n'=1,2,...n. And the maxima of the probability density are given by
$$g_{max} \cong n' \alpha_g^2 / (2n^3 R) \tag{38}$$
Because of the large factor $n^3$ in the denominator this is appreciable only for n'=n Thus we take n'=n in Eq.(37) and (38) to get
$$r_n \cong (n^2 R/\alpha_g^2)(\pi^2/8) \tag{39}$$
It is interesting to note that the value given by the semiclassical Bohr theory is
$$r_n \cong n^2 R/\alpha_g^2 \tag{40}$$
Since $\pi^2/8$ in Eq. (39) is of the order of unity the two results, Eqs. (39) and (40), are comparable. Since the area of a black hole never decreases and since the black holes in a stable holeum must not overlap, all bound state radii $r_n$ must exceed twice the black hole horizon radius in appropriate coordinates. Naively one might be tempted to say $r_n > 2R$. Strictly speaking we are prevented from exracting such a precise inequality, however, by the fact that our analysis is only valid for values of r >> R.  A true "nonoverlap" condition for the black holes can only really be extracted in the strong field regime, where our analysis is not valid. In the strong field case care must be taken, since the position of the black hole horizon is coordinate-dependent, and in terms of an isotropic radial coordinate for which $r^2 = x^2 + y^2 + z^2$ in terms of asymptotically Euclidean coordinates x,y,z (and which ,therefore is the radial coordinate appropriate to the weak field limit used here),the position of the horizon is actually at r = R/4. However, black holes which are very close together can really only be analysed by a full general relativistic solution to the appropriate two-body problem. Coordinate distances can not be expected to be simply additive in this case. We nonetheless find that it is useful to extract a rough dividing line between stable and unstable holeums. We will, therefore, define the gravitational radius of a black hole in a *purely Newtonian sense* as the radius for which the escape velocity from a spherical body of mass m is equal to the velocity of light. This singles out the Schwarzschild radius, Eq.(31),as the "Newtonian black hole radius". As outlined above, using 2R as the minimum possible separation of Newtonian "black holes" we are led to conclude
$$r_n > 2R \tag{41}$$
for all n and at all times. Now we consider two cases:

$$\alpha_g^2 < \pi^2/16 \tag{42}$$
and
$$\alpha_g^2 > \pi^2/16 \tag{43}$$
From Eqs. (39) and (42) we see that the condition for nonoverlap, Eq. (41), can be satisfied for all n, including n=1, the ground state. In this case the holeum is as stable as a Hydrogen atom. On the other hand when $\alpha_g$ is given by Eq. (43) the nonoverlap condition, Eq. (41) can be satisfied only if
$$n^2 > 16\alpha_g^2/\pi^2 > 1 \tag{44}$$

In this case the holeum can exist only in excited states and the system is denied the stable ground state n=1. This will eventually result in the coalescence of the constituent black holes and the destruction of the holeum as discussed in reference [1].

Now we would like to discuss the validity of Eq.(39). It is derived in the framework of NG which applies for r>>R. Now a holeum of atomic size has a ground state radius of about $10^{-10}$ m whereas R< $10^{-35}$ m. Thus, the condition r>>R is eminently satisfied. Similar considerations apply to the nuclear-sized holeum. In fact, even if we take the black hole mass as big as m = 0.1 $m_P$, we get the holeum radius

$$r_n = 10^4 n^2 R(\pi^2/8) \tag{45}$$

And this, too, eminently satisfies the NG condition, $r_n$ >>R. This means that all the results presented in the Table II, except the first one, are exact. Now consider the dividing line

$$\alpha_g^2 = \pi^2/16 \tag{46}$$

between the unstable and the stable holeums. This corresponds to the mass of the black holes m = 0.8862$m_P$. For this case the NG breaks down as expected from our discussion above. Nevertheless, we note that the potential for r>>R is still $r^{-1}$ and in view of the BSW and the POTHA presented in the Introduction, we might still hope to get reasonable order of magnitude values of the bound state parameters in this strong field regime. In summary, Eq.(39) is to be regarded as an asymptotic expression in the strong field case and exact elsewhere, whereas the inequalities in Eqs.(41)-(44) are to be regarded as probably only valid to within an order of magnitude. This caveat must be borne in mind in what follows.

The mass of the bound state is given by

$$M_n = 2m + E_n/c^2 \tag{47}$$

Substituting Eq. (27) into Eq. (47) we get

$$M_n = 2m(1 - \alpha_g^2/(8n^2)) \tag{48}$$

From Eqs. (30), (39) and (48) we get

$$M_n/r_n = (16 \alpha_g^2/(\pi^2 n^2))(c^2/2G)(1 - \alpha_g^2/(8n^2)) \tag{49}$$

In view of Eq. (42), we see that for a stable holeum

$$M_n/r_n < c^2/2G \tag{50}$$

for all n. This shows that a stable holeum satisfying Eq. (42) is not a black hole. With the help of Eq. (25) we can rewrite the condition for a stable holeum, Eq. (42) as.

$$m < (\pi^{1/2}/2) m_p \equiv m_c \tag{51}$$

where $m_c$ will be called the cosmic limit for the formation of a stable holeum. The numerical value of $m_c$, Eq. (51), is

$$m_c = 0.8862 \, m_p \tag{52}$$

whereas the semiclassical Bohr result, Eq. (40), gives a slightly different value $m_c = 2^{-1/4} m_p = 0.8409 \, m_p$. Thus, we find that if each of the masses of two identical black holes is less than $m_c$ then they will form a stable holeum. We note that if the black holes have unequal masses $m_1$ and $m_2$ then the condition for a stable holeum would be

$$(m_1 m_2)^{1/2} < m_c \tag{53}$$

Eq. (51) is both the necessary and the sufficient condition for the nonoverlap of the constituent black holes of a stable holeum embodied in Eq. (41). Not only that, it guarantees that the holeum will be stable. Eq.(51) implies Eq.(41) which, in turn, implies that a holeum occupies space just like ordinary matter. Its size can not be reduced below 2R. This nonoverlap property is similar to the Pauli exclusion principle and reminds us of the following result from the second quantised field theory. If we try to second-quantise a spinor field using a commutation rule rather than an anticommutation one, then there is no lower bound on the energy of the bound state and there will be no stable fermions in the universe. On the other hand, if we quantise a spinor field using an anticommutation rule then there is a lower bound and the system is stable. The anticommutation rule leads to the exclusion property. This is a spin-dependent property. In our case we have derived Eq.(39) from the maxima of the probability density which has no classical analogue. This is a purely quantum mechanical property except that it is a mass-dependent one. If the mass of the black holes is less than $m_c$, there is no overlap and the system also has the ground state n= 1. If the mass is greater than $m_c$, they overlap and the ground state n = 1 is unavailable to them. They would annihilate.

## III  GRAVITATONAL RADIATION :

If a holeum in an excited state n' goes to a lower energy state n, it emits a gravitational radiation of frequency given by

$$\nu = \nu_g (n^{-2} - n'^{-2}) \tag{54}$$

where

$$\nu_g = mc^2 \alpha_g^2/4h \tag{55}$$

which is the gravitational Rydberg constant. Note that the orbital angular momentum must change by a multiple of 2 because of the spin 2 of the graviton. Now $n = n_r + l + 1$ where $n_r$ is the radial quantum number and l is the orbital angular momentum quantum number. Therefore n' – n must be an even integer. Note that we can rewrite Eq. (55) using Eq. (25) as

$$\nu_g = \nu_0 (m/m_p)^5 \tag{56}$$

where

$$\nu_0 = m_p c^2/4h = 7.416346 \cdot 10^{41} \text{ Hz} \tag{57}$$

and

$$r_n = n^2 (\pi^2/8)(R_p/R)^3 R_p = R_g n^2 \tag{58}$$

where $R_p$ is the Schwarzschild radius of a black hole of mass $m_p$. It is given by.

$$R_p = 2m_p G/c^2 = 1.609502 \cdot 10^{-35} \text{ m} \tag{59}$$

and

$$R_g = (\pi^2/8)(R_p/R)^3 R_p \tag{60}$$

Using Eqs. (56) – (59), we present the values of several quantities of interest in Table II. We note that presently gravitational wave detectors are being built around frequencies of about $10^2$ to $10^3$ Hz. From the Table II we see that holeums of sizes lying between the atomic and the nuclear ones would emit gravitational radiation covering this range. This can be tested once sufficiently powerful gravitational wave detectors become operational. The corresponding masses of the primordial black holes are in the range $10^{11}$ GeV/$c^2$ to $10^{12}$ GeV/$c^2$ as seen in the Table II.

From Eq. (58) we get

$$dr_n/dn = 2nR_g \tag{61}$$

This quantity, $R_g$ given by Eq. (60), is presented in the last column of Table II. From the latter we see that $R_g$ is extremely small for massive holeums. This makes the derivative in Eq. (61) negligibly small implying great stability of their orbits. As seen from Eqs. (56) and (57), the transitions among the discrete levels n<10, say, of the massive holeums are very costly in terms of the energy. And, as discussed in reference 1, the levels with n>>1 have large orbital angular momenta. Hence the transitions among them are very costly in terms of the angular momentum. The overall result is that the massive holeums have much more stable orbits than the lighter ones. Hence the lighter ones are more likely to radiate gravitational waves than the heavier ones. And from the Table II we see that the lighter holeums emit frequencies in the detectable range.

## IV  DARK MATTER AND DOMAIN WALLS :

A holeum is a gravitational analogue of a hydrogen atom as regards its energy spectrum. Nature would seem to have produced two types of hydrogen atoms in the early universe: electromagnetic and gravitational. The former is well known but the latter remains to be discovered because of our lack of gravitational wave detectors. Now particles of ordinary matter have all the four interactions whereas the holeums have only the gravitation, the weakest. Hence they will tend to segregate from the former. And still the holeums will cling to the galaxies and other formations of ordinary matter in the form of haloes because of gravity. These haloes remain undetected because of our lack of gravitational wave detectors. Because of copious production of black holes in the early universe, the holeums would make up an important component of the dark matter in the universe today.

One implication of the Standard Model is that the universe may have had a number of ground states separated by domain walls [12]. No such walls have been detected. This is one of the open questions of the standard model of cosmology. It is tantalizing to speculate that the domain walls may consist of haloes of the holeums. Then their nonobservation at present would be obvious.

## V  TERRESTRIAL AND ASTROPRODUCTION OF HOLEUMS

In Proton – antiproton and other particle – antiparticle collisions in high energy accelerators mini-bigbang-like highly excited states of vacuum are created. To see if holeums can be produced at the currently available accelerator energies we look at Table II. We find that two black holes of mass 1 TeV/$c^2$ each, at the currently available energies, will form a holeum of radius $10^{15}$ m which is much larger than the size of the interaction region in an accelerator. Thus, production of holeums in the present accelerators is ruled out. Again from the Table II we see that to produce a holeum of radius 1 $^0$A, two black holes of masses of the order of $10^{10}$ GeV/$c^2$ at a temperature much grater than $10^{23}$ $^o$K will have to be produced. This is far beyond the range of any accelerator in the foreseeable future. The highest temperature in Tokamak plasmas at present is of the order of 1 KeV or $10^7$ $^o$K. This is far too low for the production of holeums in Tokamaks for a long time to come. Neither in the explosions of thermonuclear devices of megatonne variety nor in the interiors of stars do we have temperatures required to produce holeums. If temperatures of the order of $10^{25}$ $^o$K are produced in supernova explosions, then holeums may be produced in there. Barring this, we will have only the primordial holeums.

## VI CONCLUSIONS

We have presented three arguments to make it plausible that gravitational bound states of black holes may be produced in the early universe. We also find a purely quantum mechanical exclusion principle dependent upon the mass of the black holes. If the mass is less than $m_c$, given by Eqs.(51) and (52), then the two black holes would not overlap each other and a stable holeum results, not otherwise. We must note, of course, that Eqs.(51) and (52) involve a bound using entirely Newtonian arguments in the strong field regime of gravity and so the exact value of $m_c$ can not be relied upon unless it turns out to be rigorously justified from GR. However, we expect that in the GR case a similar bound would exist, and that it would give a value close to that of Eq.(51) to within perhaps an order of magnitude.   The holeums are not black holes. They occupy space. The transitions among the energy levels of a holeum result in the emission of gravitational waves whose frequencies are predicted. Their gravitational spectra are formally identical with the electromagnetic spectra of a hydrogen atom. This is the most important prediction of this model. A part of their spectrum lies in the region accessible to the detectors being built. The holeums would be produced copiously in the early universe and therefore would form an important component of the dark matter in the present day universe. The holeums would form haloes around the galaxies and other concentrations of the ordinary matter in the universe. They may also be the constituents of the invisible domain walls predicted by the Standard Model of cosmology.

TABLE I.

| $\varepsilon$ | p | $\lambda_1$ | $\lambda_2$ | $E_0$ eV | $r_0$ A | $r_{max}$ A | $r_{max}/a$ |
|---|---|---|---|---|---|---|---|
| 0.3* | 1.072190 | 1.206078 | 0.830515 | -8.16 | 0.55 | 1.65 | 3.23 |
| 0.307922 | 1.0* | 1.176502 | 0.784758 | -8.38 | 0.51 | 1.60 | 3.13 |
| 0.4* | 0.387247 | 0.809028 | 0.346364 | -10.88 | 0.20 | 0.09 | 0.18 |

TABLE II

| m<br>GeV/c$^2$ | $\alpha_g$ | $T_b$<br>$^0$K | $\nu_g$<br>Hz | $R_g$<br>m |
|---|---|---|---|---|
| 1.23 E 19 | 1.00 | 1.43 E 32 | 7.42 E 41 | 3.98 E −35 |
| 1.23 E 17 | 1.00 E- 04 | 1.43 E 30 | 7.42 E 31 | 3.98 E −29 |
| 1.23 E 15 | 1.00 E -08 | 1.43 E 28 | 7.42 E 21 | 3.98 E −23 |
| 1.23 E 13 | 1.00 E -12 | 1.43 E 26 | 7.42 E 11 | 3.98 E −17 |
| 1.82 E 12 | 2.18 E -14 | 2.11 E 25 | 5.23 E 07 | 1.23 E -14 |
| 3.91 E 11 | 1.01 E -15 | 4.54 E 24 | 2.43 E 04 | 1.23 E -12 |
| 8.43 E 10 | 4.71 E -17 | 9.79 E 23 | 1.13 E 01 | 1.23 E -10 |
| 1.82 E 10 | 2.18 E -18 | 2.11 E 23 | 5.23 E -03 | 1.23 E -08 |
| 3.91 E 09 | 1.01 E -19 | 4.54 E 22 | 2.43 E -06 | 1.23 E -06 |
| 8.43 E 08 | 4.71 E -21 | 9.79 E 21 | 1.13 E -09 | 1.23 E -04 |
| 1.82 E 08 | 2.18 E -22 | 2.11 E 20 | 5.23 E -13 | 1.23 E -02 |
| 1.23 E 03 | 1.00E -32 | 1.43 E 16 | 7.42 E -39 | 3.98 E 15 |

TABLE CAPTIONS

Table I. Parameters of Truncated Hydrogen Atom . $r_{max}$ is the most probable radius. Starred quantities are the inputs.
TableII. Values of some parameters of holeums . See the text for definitions.